\documentclass[12pt]{iopart}

\usepackage{graphicx}
\usepackage{bm}

\begin{document}

\title{Separation of particles leading either to decay and unlimited growth of
energy in a driven stadium-like billiard}

\author{Andr\'e L.\ P.\ Livorati$^{1,2}$, Matheus S. Palmero $^{2,3}$, Carl P.
Dettmann$^{2}$, Iber\^e L. Caldas $^{1}$ and Edson D.\ Leonel$^{3}$}

\address{$^1$Instituto de F\'isica - IFUSP - Universidade de S\~ao Paulo -
USP  Rua do Mat\~ao, Tr.R 187 - Cidade Universit\'aria - 05314-970 - S\~ao
Paulo - SP - Brazil\\
$^2$ School of Mathematics - University of Bristol - Bristol BS8 1TW - United
Kingdom\\ 
$^3$ Departamento de F\'isica -- UNESP -- Universidade Estadual Paulista -- Av. 24A,
1515 - Bela Vista - 13506-900 - Rio Claro - SP - Brazil
}

\begin{abstract}
A competition between decay and growth of energy in a time-dependent stadium
billiard is discussed giving emphasis in the decay of energy mechanism. A critical resonance velocity
is identified for causing of separation between ensembles of high and low
energy and a statistical investigation is made using ensembles of initial
conditions both above and below the resonance velocity. For high initial
velocity, Fermi acceleration is inherent in the system. However for low
initial velocity, the resonance allies with stickiness hold the particles in a
regular or quasi-regular regime near the fixed points, preventing them from
exhibiting Fermi acceleration. Also, a transport analysis along the velocity
axis is discussed to quantify the competition of growth and decay of energy
and making use distributions of histograms of frequency, and we set that the causes
of the decay of energy are due to the capture of the orbits by the resonant fixed points.

\end{abstract}

\maketitle

\section{Introduction}
\label{sec1}
~~~~~Modelling dynamical systems with mixed phase space has been 
one of the main challenges of the nonlinear statistical mechanics research field and has received especial
attention in the last decades \cite{ref1,ref2,ref3,ref4}. Particularly, with the advance of fast
computers the dynamics can be evolved over long time series letting several
phenomena, some of them completely new to be observed.
A class of dynamics of particular interest includes the Hamiltonian 
systems with time-dependent perturbation, for which energy varies with time.
Moreover, a better understanding of the mixed dynamics in the phase space can be given, and
phenomena related to them carefully characterized. The study of chaotic properties of systems 
can be found in many fields  of physics such as fluids \cite{ref5}, plasmas \cite{ref6,ref7}, 
nanotubes \cite{ref8}, complex networks \cite{ref9}. 
Interesting applications and phenomena can be found in optics \cite{ref10,ref11,ref11a} and acoustic \cite{ref12} 
if billiard dynamics is considered. Particularly, if the billiard boundaries are time dependent, applications in   
microwaves, \cite{ref12a,ref12b,ref12c} and quantum dots \cite{ref12d,ref12e,ref12f,ref12g} can also be found.
Also, when a particle-particle iteration in billiard dynamics is considered, one could find synchronization \cite{n1,n2} 
and soft wall effects \cite{n3,n4}.

In particular, a phenomenon which intrigues physicists in general is the
unlimited energy growth of a bouncing particle with a driven boundary, a
phenomenon called Fermi Acceleration (FA). It was introduced in earlier
1949 by Enrico Fermi \cite{ref13} as an attempt to describe the mechanism in
which particles that have their origin from cosmic rays, acquire very high
energy. Nowadays his idea is extended to other models where average properties
of the velocity or kinetic energy, for long time series, are studied.
In light of this approach, a billiard type dynamics is one of the most useful
systems \cite{ref14} which possibly exhibits unlimited energy growth when the
boundary is moving in time. The Loskutov-Ryabov-Akinshin (LRA) conjecture
then gives the
minimal conditions to observe such a phenomenon \cite{ref15,ref16}. It then
claims that if a billiard has a chaotic component for the dynamics in the
static boundary case, this is a sufficient condition to observe unlimited
energy growth when a time-dependence on the boundary is introduced.
The elliptical billiard is not described by the LRA conjecture. It is indeed
integrable for the static boundary case presenting a separatrix curve
separating two types of dynamical regimes: libration from rotation. The
introduction of time perturbation in the boundary makes the separatrix curve
to turn into a stochastic layer letting the particle suffer successive
crossings from rotation to libration regions \cite{ref17,ref18,ref19} leading
to an unlimited diffusion in energy.

Recent study \cite{ref20}, indicates that non-linear phenomena like
stickiness can act as a slowing mechanism of FA. In fact, the finite time
trappings around the stability islands influence some transport properties,
making the system locally less-chaotic \cite{n5}. Such influence of stickiness can be
found in several models and applications on the literature
\cite{ref3,ref4,ref20a,ref21,ref22,ref23,ref24}. In this paper we revisit the
problem of a stadium-like billiard with oscillating circle arcs boundaries,
focusing on the analysis of the mechanism that produces the
decay of energy, where many low energy orbits are observed to undergo reduction of energy to
an apparently stable state. We argue for existence of a critical
resonance velocity where high initial velocities produce FA and low initial
velocities do not experience the unlimited energy grow. Such decay is caused
by influence of sticky orbits with resonance around the stability
islands. It leads the chaotic orbits to suffer a temporary trapping around
stability islands and then be captured by the fixed points. A statistical investigation
is made in order to quantify this phenomenon. Similar approach was done previously
\cite{ref25,ref26,ref27,ref28,ref29,ref30}, but in this paper we emphasize the decay of
energy and its origin in stickiness for the first time. We also discuss the statistics of
transport in the velocity axis near a resonant velocity marking a
separation of low energy to high energy regimes producing the decay of energy
and the phenomenon of unlimited energy growth.

The paper is organized as follows. In Sec.\ref{sec2} we describe the dynamics
of the stadium billiard and a study of its chaotic properties. Section
\ref{sec3} is devoted to discuss the statistical and transport analysis of the
properties for both ensembles of initial conditions considering low and high
energy regime. Here we also present the influence of the stickiness orbits
that hold the orbits in a quasi periodic motion near the fixed points, what
causes the decay of energy. Finally in Sec. \ref{sec4} our final remarks and
conclusions are presented.

\section{Stadium billiard as model, the mapping and chaotic properties}
\label{sec2}

~~~~~This section is devoted to discuss the model and the equations describing
the dynamics. The model consists of a classical particle (or an ensemble of
non-interacting particles) moving inside a closed domain of a stadium-like
shape. The stadium billiard is composed by two parallel lines connected with
regions of negative curvature \cite{ref31,ref31a}. In this paper we consider the
boundary of the stadium described by three geometric control
parameters, $a$ which is the width of the circle arc, $b$ indicates the depth 
of the curvature and $l$ is the strength of the parallel lines. 
Additionally, we introduce in the boundary a time dependence. The
dynamics of a particle for the static version of the billiard are characterized
by a constancy in the energy. However the defocusing mechanism, as proposed
originally by Bunimovich \cite{ref31}, is responsible to generate chaotic
dynamics under the condition $(4bl/a^2)\approx(l/2R)>1$ \cite{ref31a,ref32}, where $R$ is the original
radius of the Bunimovich stadium. According to the LRA conjecture, a chaotic dynamics is a sufficient
condition to produce unlimited energy growth in the velocity of the particle
when a time perturbation to the boundary is introduced. The unlimited
diffusion in velocity generated due to collisions of a particle with a massive
and time moving boundary is known in the literature as Fermi acceleration
\cite{ref13}. Robustness is not a characteristic of the phenomenon since
inelastic collision \cite{add1,add2,add2a,add3} as well as dissipation introduced by
the drag-type force \cite{add4,add5} suppresses the unlimited diffusion.

\begin{figure}[ht!]
\centering
\includegraphics[width=7cm]{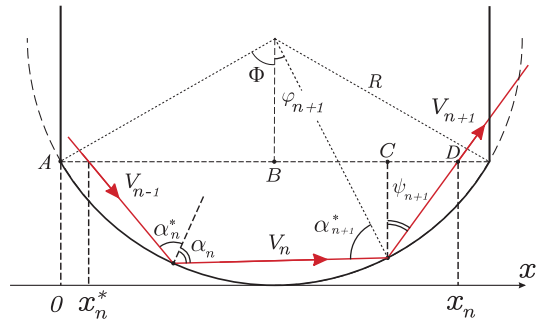}
\caption{{\it Sketch of the model near a focusing boundary and illustration
of a typical trajectory suffering successive collisions.}}
\label{fig1}  
\end{figure}

\begin{figure}[ht!]
\centering
\includegraphics[width=7cm]{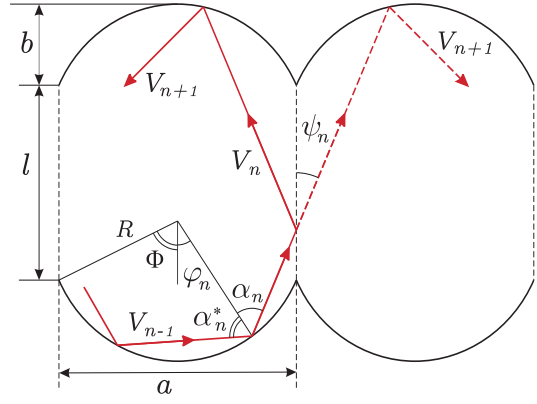}
\caption{{\it An illustration of the indirect collisions dynamics and its variables and parameters.}}
\label{fig2}  
\end{figure}

As usually made in the literature, the dynamics of the particle are made by
using a nonlinear mapping. We therefore take into account two distinct
possibilities of the dynamics which include: (i) successive collisions and;
(ii) indirect collisions. For case (i), the particle suffers successive
collisions with the same focusing component while in (ii), after suffering a
collision with a focusing boundary, the next collision of the particle is with
the opposite focusing boundary. A schematic illustration of the both collision
cases can be found in Figs. \ref{fig1}, \ref{fig2}. In between such a collision the particle can,
in principle, collide many times with the two parallel borders. We have
considered that the time dependence in the boundary is $R(t)=R_0+r\sin(wt)$,
where $R_0=(a^2+4b^2)/8b$ is the radius of the static boundary and  $R_0\gg
r$. The definition of the parameters $a$ and $b$ can be found in Fig. \ref{fig2}. 
The velocity of the boundary is obtained by
\begin{equation}
\dot{R}(t)={B(t)}=B_0\cos(wt)~,
\label{eq1}
\end{equation}
with $B_0=rw$ and $r$ is the amplitude of oscillation of the moving boundary
while $w$ is the frequency of oscillation. In our investigation however we
consider only the so called static boundary approximation (also called as a
simplified version). It assumes that the boundary is fixed, which makes the
time spent between collisions easy to be calculated. However the velocity
after collision is calculated as if the boundary were moving. For this 
kind of approximation, we may find some examples of dynamical systems, 
whose behaviour is basically the same, considering a comparison between simplified 
and complete version dynamics. The Bouncer model and the Standard map are two of the examples \cite{ref20,add2,add2a},
and the stadium billiard itself \cite{ref27,ref28}, shows a very similar dynamics for both versions. 

The mapping is constructed for the variables $\alpha_n$ corresponding to the
angle between the trajectory of the particle and the normal axis at the
collision point, $\varphi_n$ is the angle measured between the normal axis at
the collision points and the symmetry line of the vertical axis of the
stadium, $t_n$ denotes the time and $V_n$ the outgoing speed of the particle 
for the nth collision. The condition to observe case (i),
i.e. the successive collisions, is $|\varphi_{n+1}|<\Phi$, where
$\Phi=\arcsin(a/2R_0)$ is the angle between the vertical symmetry line and the
maximum angle of the negative curvature region. A typical successive
collision case is represented in Fig. \ref{fig1}. Using basic geometry
properties we obtain
\begin{equation}
\left\{\begin{array}{ll}
\alpha^{*}_{n+1} = \alpha_n~\\
\varphi_{n+1} = \varphi_n +\pi -2\alpha_n~~ (mod~ 2\pi)~\\
t_{n+1} = t_n + {2R\cos(\alpha_n)\over V_n}~\\
\end{array}
\right.,
\label{eq2}
\end{equation}
where the superscript $(*)$ represents the dynamical variable immediately
before the collision.

Case (ii) is considered when $|\varphi_{n+1}|>\Phi$ and the particle collides
with the opposite focusing component. In principle, the particle can suffer 
several collisions with the straight walls, so for such a type of collision, we make
use of the unfolding method \cite{ref14,ref32} to describe the dynamics. Two
auxiliary variables are then introduced, $\psi$, which is the angle between
the vertical line at the collision point and the particle's trajectory, and
$x_n$, representing the projection over the horizontal axis. A sketch of the
indirect collisions is shown in Fig. \ref{fig2},
therefore leading to the following mapping
\begin{equation}
\left\{\begin{array}{ll}
\psi_n=\alpha_n-\varphi_n~~mod(\pi/2)\\
x_n={R\over\cos(\psi_n)}[\sin(\alpha_n)+\sin(\Phi-\psi_n)]~\\
x_{n+1}=x_n+l\tan(\psi_n)~~mod(a)\\
\alpha^*_{n+1}=\arcsin[\sin(\psi_n+\Phi)-x_{n+1}\cos(\psi_n)/R]~\\
\varphi_{n+1}=\psi_n-\alpha^*_{n+1}~\\
t_{n+1}=t_n+{R[\cos(\varphi_n)+\cos(\varphi_{n+1})-2\cos(\Phi)]+l\over
V_n\cos(\psi_n)}~\\
\end{array}
\right..
\label{eq3}
\end{equation}

The expression for the velocity of the particle after collision is obtained
by decomposing it into two separate components, which are
\begin{equation}
\left\{\begin{array}{ll}
\vec{v}_n \cdot \vec{T}_n = v_n\sin(\alpha^{*}_n)~~\\
\vec{v}_n \cdot \vec{N}_n = -v_n\cos(\alpha^{*}_n)~~\\
\end{array}
\right.,
\label{eq4}
\end{equation}
where $\vec{T}$ and $\vec{N}$ are the tangent and normal unit vectors at the
collision point. Because the collision is happening in a moving referential
frame (non inertial) we have to make change of referential frames from
inertial to non-inertial. The reflection law is then given by
\begin{equation}
\left\{\begin{array}{ll}
\vec{V}^{\prime}_{n+1}\cdot\vec{N}_{n+1}=-\kappa\vec{V}^{\prime}_n\cdot
\vec{N}_{n+1}~~\\
\vec{V}^{\prime}_{n+1}\cdot\vec{T}_{n+1}=\eta\vec{V}^{\prime}_n\cdot
\vec{T}_{n+1}~~\\
\end{array}
\right.,
\label{eq5}
\end{equation}
where $\kappa\in[0,1]$ and $\eta\in[0,1]$ are the respective restitution
coefficients for the normal and the tangent components, and the superscript
$^{\prime}$ resembles the non-inertial referential frame. In this paper we
consider only the conservative case, therefore $\kappa=\eta=1$, although the
construction of the mapping is more general.

Moving back to the inertial referential frame, the components of the
velocity of the particle after collision are given by
\begin{equation}
\left\{\begin{array}{ll}
\vec{V}_{n+1}\cdot\vec{N}_{n+1}=-\kappa
\vec{V}_n\cdot\vec{N}_{n+1}+(1+\kappa)\vec{B}(t_{n+1})\cdot\vec{N}_{n+1}~~\\ 
\vec{V}_{n+1}\cdot\vec{T}_{n+1}=\eta
\vec{V}_n\cdot\vec{T}_{n+1}+(1-\eta)\vec{B}(t_{n+1})\cdot\vec{T}_{n+1}~~\\ 
\end{array}
\right.
\label{eq6}
\end{equation}
Finally the velocity of the particle after collision is given by
\begin{equation}
\mid\vec{V}_{n+1}\mid=\sqrt{(\vec{V}_{n+1}\cdot\vec{T}_{n+1})^2+(\vec{V}_{n+1}
\cdot\vec{N}_{n+1})^2}~.
\label{eq7}
\end{equation}

The two components of the velocity can be used to obtain the angle
$\alpha$, leading to
\begin{equation}
\alpha_n=\arcsin\left[{{{\mid\vec{V}_n\mid}\over{\mid\vec{V}_{n+1}\mid}}
\sin(\alpha^{*}_n)}\right]~.
\label{eq8}
\end{equation} 
The dynamics of the particle is then evolved by considering simultaneously
equations (\ref{eq2}), (\ref{eq3}), (\ref{eq7}) and (\ref{eq8}).

\section{Numerical Results and Statistical Investigation}
\label{sec3}

~~~~~In this section we present and discuss on how the stickiness orbits influence
the dynamics. Moreover we proceed the study by doing an extensive statistical
investigation in the dynamics particularly considering distributions of the
angular variables along the dynamics. The novelty here is that after
considering the histograms of frequency analysis for either the velocities and
the polar angles for the decay of the velocity for a very long time series, we
can see that the orbits after experience the stickiness influence, are
captured by the fixed points, as if the dynamics was under the regime of
dissipation. This makes the behaviour of the average velocity curves decays
for lower energy ensembles. Also, a transport investigation along the velocity
axis considering both the low and high energy regimes in order to quantify
the competition between the FA and the decay of energy near the critical
resonance velocity.

\subsection{Resonance Velocity}
\begin{figure}[ht!]
\centering
\includegraphics[width=9cm]{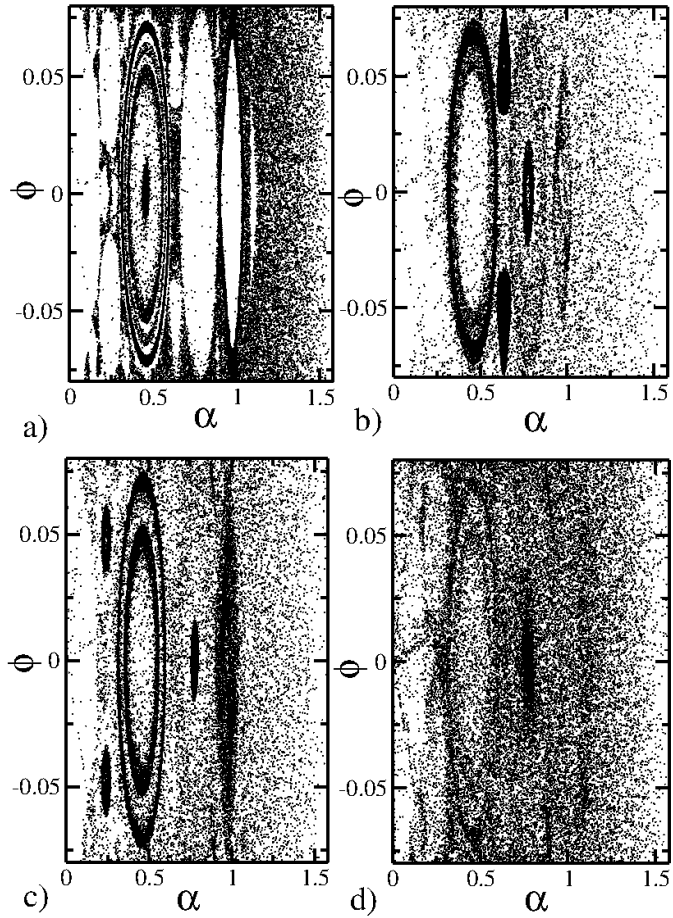}
\caption{{\it Phase space for the time-dependent stadium billiard. The initial
velocities used were (a) $V_0=5$, (b) $V_0=1.5$, (c) $V_0=1.2$ and (d)
$V_0=0.5$.}}
\label{fig3}
\end{figure}

~~~~~Let us start discussing the resonance velocity. The phase space may be represented, 
according to the convenience, either using angular coordinates $\alpha_n$ and $\phi_n$,
or the auxiliary variables $\psi$ and $\xi$, where $\xi=0.5+R_0\sin(\phi_{n+1})/a$ is
the projection along the horizontal axis, and is usually normalized at $mod(1)$.
The fixed points $\psi^{*}$ are deeply connected with $\alpha_n$ because of
the axial symmetry of the billiard. There is a sequence of fixed points
corresponding to orbits always within the stadium at $\phi_n=0$ (period-1),
that intersects different multiples of the parameter $a$ in the horizontal
direction, according to the unfolding method \cite{ref14,ref32}. They are represented in 
the phase space by the elliptical stability islands in Fig. \ref{fig3}.
Also, such stable orbits, can be understood as librator-like trajectories inside 
the billiard (see Ref. \cite{ref14} for details).

Considering the linearisation of the unperturbed mapping \cite{ref25,ref26,ref27,ref28} 
around the fixed points and according to the action-angle variables,
one finds the rotation number $\sigma=\arccos\left(1-{8bl\over
a^2\cos^2(\psi^{*}_n)}\right)$, where $\psi^{*}_n=\arctan(ma/l)$ is the fixed point and
$m\ge 1$ is the number of mirrored {\it stadiums} that the particle can go
through in a trajectory (for detailed explanation see Ref. \cite{ref26}).

Now, considering an orbit of a particle moving around a fixed point in the
unperturbed (static) version of the billiard, the time spent between two
successive collisions, i.e., the time between collisions with the two
focusing components is $\tau \approx {l\over\cos(\psi^{*}_n)V_n}$ 
(considering the period-1 fixed point). Thus the rotation period of such 
an orbit around the fixed point $T_{rot}={2\pi\tau\over\sigma}$.

When a time perturbation is introduced in the focusing boundaries,
there is an external perturbation period given by $T_{ext}={2\pi\over w}$.
When $T_{rot}=T_{ext}$ there is then a resonance between the moving boundaries
oscillations and the rotating orbits around the fixed points (see Ref. \cite{ref26} 
for explanation figures about this resonance). After grouping
the terms properly one finds
\begin{equation}
V_r={wl\over\cos(\psi^{*}_n)\arccos\left(1-{8bl\over(a\cos(\psi^{*}_n))^2}
\right)}~.
\label{eq9}
\end{equation}

Of course, that the value of $V_r$ depends on what fixed point we are evaluating the 
linearisation. However, for all period-1 fixed points, one may found a very close value
for the resonant velocity for all of them. For the combination of the parameters as
$l=1.0$, $a=0.5$, $b=0.01$, we have $V_r=1.2$ which is an average of all possible values 
for the different period-1 fixed points that one may found with the 
combination of the geometric control parameters given below. So, around $V_r$,
all the fixed points become resonant, and increase the mixing in the phase space.
For a better understanding of the influence of these parameters on the number 
of islands and fixed points in the phase space and their relation 
with the defocusing mechanism please see Ref. \cite{ref31a,ref32}. Also, it
must be stressed that such resonance is only observed when the defocusing
mechanism is no longer active \cite{ref31a,ref32}, once with fully chaotic phase space, 
no stability island exists, therefore no resonance is observed.

Indeed resonance is a phenomenon often observed in dynamical systems with
mixed phase space properties. When an orbit has its velocity equal or less
than the resonant one, the particle can penetrate in the neighbourhood of the
fixed points. It may then enter the stability island for a while and leave
such a region \cite{ref2,ref3,ref4,ref20a} or may be trapped in a pseudo stable orbit
for a long time. Such behaviour is like it was attracted by the fixed point (we
show this in the next sections). A possible explanation for this kind of
phenomenon is related to a transformation of an invariant curve observed in
the static version into a porous curve as the time perturbation is considered.
Then a particle may cross a porous curve making possible visits to previous
forbidden regions which become now accessible for the particle.

Figure \ref{fig3} shows a phase space for different initial velocities
considering both $V_0<V_r$ and $V_0>V_r$ for 25 different initial conditions
chosen along the chaotic sea. As mentioned before, the phase space may be
represented,
according to the convenience, either using angular coordinates $\alpha_n$ and $\phi_n$, 
or the auxiliary variables $\psi$ and $\xi$. Figure \ref{fig3} was constructed using 
$\alpha$ and $\phi$ for $\alpha\in[0,\pi/2]$ and $\phi\in[-\Phi,+\Phi]$. 
Stickiness regions can be seen in Fig. \ref{fig3} particularly when the
initial velocities are given close but still below to the resonant velocity.
As the dynamics evolves, the orbits often change from an island to a
surrounding region leading to successive trappings thus not letting the
velocity of the particle to reach higher values. For long enough time the
orbit chooses (so far a proper mechanism is not yet known) an island and stay
there for really long time (as far we have studied, more then $10^9$
collisions), as if it was attracted to the fixed point into a stable orbit.
The temporally trapping around the stability islands is a possible reason to
explain the decay of energy for an ensemble of low energies $(V_0<V_r)$, where many orbits 
are observed to undergo reduction of energy to an apparently stable state. 

\begin{figure}[ht!]
\centering
\includegraphics[width=9cm]{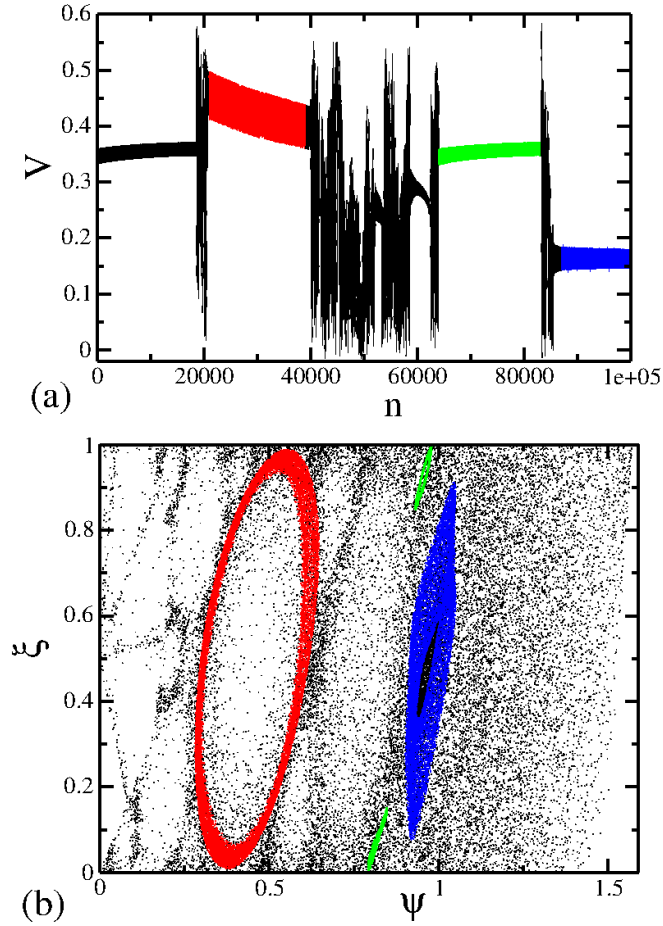}
\caption{Color Online:{\it Influence of stickiness orbits for the dynamics. In
(a) is shown the evolution of the velocity as a function of the number of
collisions, and in (b) is shown the same orbit for the coordinates $\psi$ and
$\xi$. The initial velocity considered was $V_0=0.35$.}}
\label{fig4}
\end{figure}
The mixed behaviour of the dynamics between quasi-periodic and chaotic region
is shown in Fig. \ref{fig4} where an orbit with initial velocity lower than
the resonant one is evolved for a long time leading to temporarily trappings
as shown in Figs. \ref{fig4}(a,b). A change in the coordinates of the phase
space is made from polar angles $(\alpha,\phi)$ to the auxiliary one
$(\psi,\xi)$, with $\xi=0.5+R_0\sin(\phi_{n+1})/a$.

\subsection{Statistical Analysis}  
~~~~~Let us start this section by discussing some statistical analysis concerning
behaviour of the average velocity. We consider the quadratic deviation of
the average velocity as
\begin{equation}
\omega(n,V_0)= {1\over M} \sum_{i=1}^M
\sqrt{{\overline{V_i^{2}}(n,V_0)}-{\overline{V_i}} ^ 2 (n,V_0)}~,
\label{eq10}
\end{equation}
where $M$ represents an ensemble of initial conditions. The average velocity
is given by
\begin{equation}
\overline{V}(n,V_0)= {1\over n} \sum_{i=1}^n {V_i}~.
\label{eq11}
\end{equation}

\begin{figure}[ht!]
\centering
\includegraphics[width=9cm]{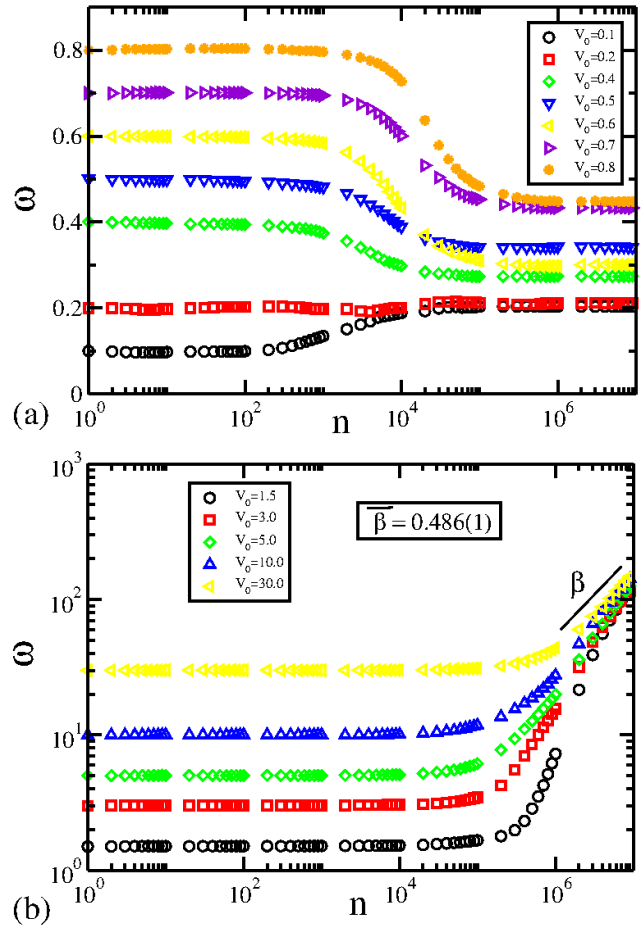}
\caption{Color Online:{\it Behaviour of $\omega$ as a function of $n$ for
ensembles of low and high initial velocities. In (a) all curves experience
a decay of energy caused by the successive stickiness trappings while in (b)
the curves are experiencing a diffusion in energy leading to Fermi
acceleration. They exhibit a growth according a power law with exponent
$\beta\approx 0.5$ for long time. The initial velocities are labelled in the
figure.}}
\label{fig5}  
\end{figure}  

\begin{figure*}[t]
\begin{center}
\centerline{\includegraphics[width=17cm,height=12.0cm]{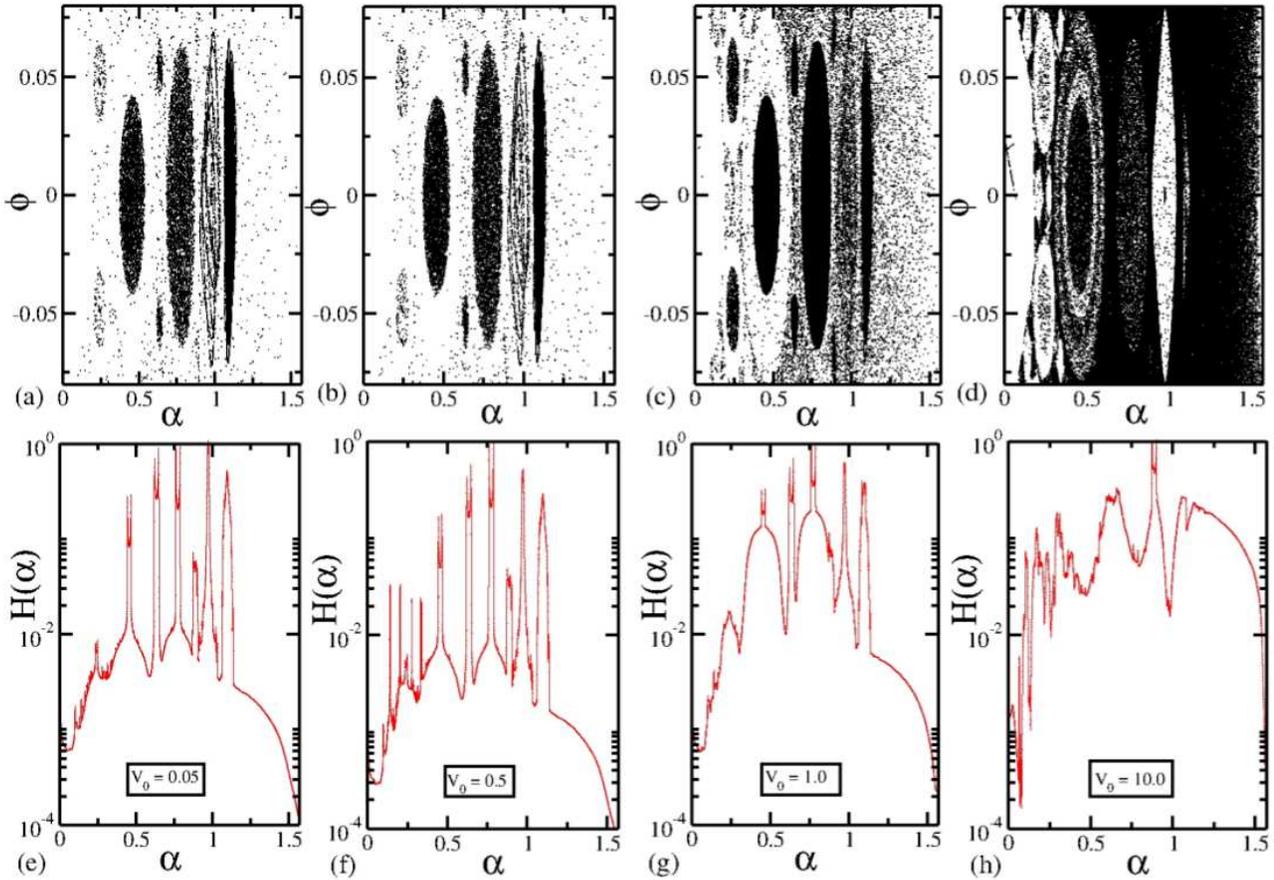}}
\end{center}
\caption{Color Online:{\it Comparison between the final pair of the angular
coordinates ($\alpha,\phi$) for both low and high energy ensembles, with the
frequency histogram along the whole dynamical evolution of the $\alpha$
coordinate. In (a) and (e) $V_0=0.05$, (b) and (f) $V_0=0.5$, (c) and (g)
$V_0=1.0$, and finally in (d) and (h) $V_0=10.0$. The histogram axis of (e),
(f), (g) and (h) are plotted in logarithmic scale.}}
\label{fig6}  
\end{figure*}

Figure \ref{fig5} shows the evolution of different curves of $\omega$ as a
function of $n$ for different initial velocities. Each curve was constructed
considering an ensemble of $2000$ different initial conditions chosen along
the chaotic sea. They were evolved in time up to $10^7$ collisions with the
boundary. One sees in Fig. \ref{fig5}(a) all the initial velocities are lower
than the resonant one ($V_0<V_r$) and the $\omega$ curves stay constant for
short times. After a crossover they experience a decay for long time series.
On the other hand, Fig. \ref{fig5}(b) shows some curves of $\omega$ for
initial velocities higher than the resonant one ($V_0>V_r$). They initially
present a constant plateau in the same range of their initial velocities and
then suddenly bend towards a growth regime marked by a power law
($\omega\propto n^\beta$) with exponent $\beta\approx 0.5$. The $\beta$ exponent on
the increments in velocity variables are well described using a central limit theorem 
(CLT) \cite{add6,add7}, so that over many time steps, the distribution of displacements
is Gaussian with a variance exactly proportional of $\sqrt{n}$ leading to an unlimited 
growth in the velocity. It is important to
emphasize the curves of $\omega$ do not depend on the control parameters $a$,
$b$ and $l$ given such parameters produce a behaviour which is scaling
invariant \cite{ref32}. Therefore only one combination of the parameters is
enough to have a tendency of the behaviour. The amplitude of the time
perturbation is assumed to be constant $B_0=0.01$ given it is also scaling
invariant \cite{ref25}.

As previously discussed and as is known in the literature, the high initial
energy ensembles, those with initial velocity higher than the resonant one,
lead to Fermi acceleration \cite{ref25,ref26}. Then we shall give a particular
attention  and focus on the low energy ensemble therefore characterizing the decay of
energy mechanism and its correlation with stickiness orbits. As shown in Fig.
\ref{fig5}, the 
separation of ensembles is seen by
considering the averages of different curves of $\omega$. Because the average
velocity is one of the observables responsible to characterize the diffusion
in energy, a statistical analysis with the average velocity for a grid of
initial conditions is a natural and good procedure to quantify the dynamics,
particularly the diffusion in energy. It let us to see what initial condition
leads to growth or decrease of velocity then defining clearly which one
produces Fermi acceleration. We than consider an investigation using the
histograms of frequency technique which let us see what region in the phase
space produce a larger stickiness, as shown in Fig. \ref{fig6}.

\begin{figure}[ht!]
\centering
\includegraphics[width=9cm]{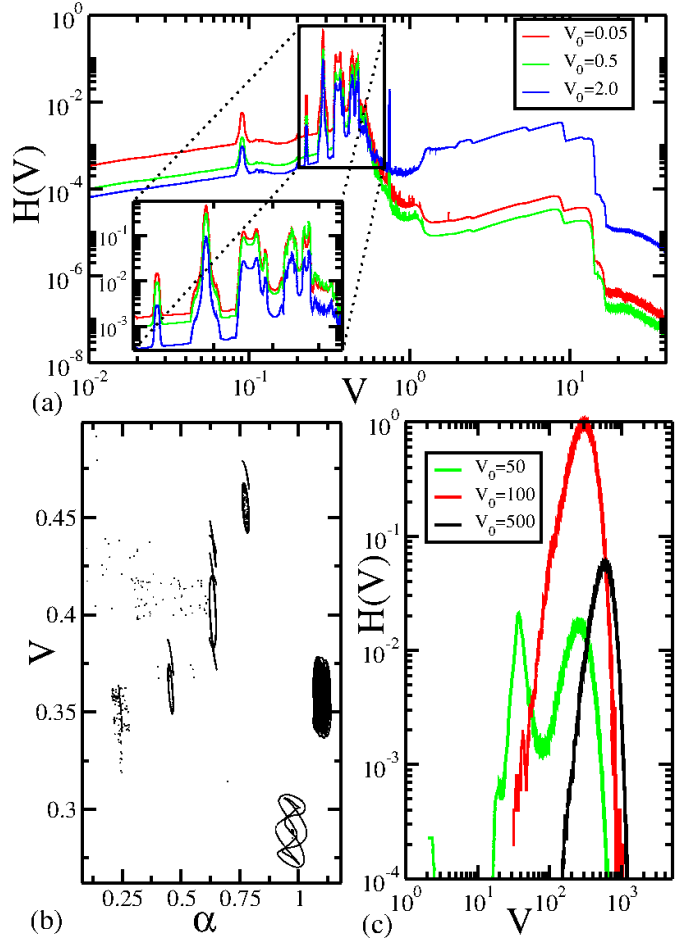}
\caption{Color Online: {\it In (a) we have the frequency histogram for the
lower energy ensemble concerning the whole velocity distribution for
$V_0=0.05$, $V_0=0.5$ and $V_0=2.0$. In particular, the zoom-window shows the
picks of more intensity. In (b) is shown the convergence velocity as function
of the $\alpha$ angle, where we are able to identify in which respective
island the orbits got trapped and with respective velocity. And in (c), we
show the frequency histogram for the high energy ensemble, where the several
peaks of intensity do not appear, and we have a very well defined
distribution. Also, the axis of (a) and (c) are in the logarithmic scale, 
for a better representation of their range and intensity.}}
\label{fig7}
\end{figure}

We start then considering the angular variables, i.e. the polar angles
($\alpha,\phi$). Since the stadium billiard has an axial symmetry, the
variable
$\phi$ (or $\xi$), is not a good choice to be considered because it leads to an almost
constant distribution along the dynamics. The same process is applied to the
auxiliary variable $\xi$. We considered a set of $5\times10^6$ initial
conditions chosen along the chaotic sea for both ensembles of high and low
energy. Each initial condition was iterated up to $10^7$ collisions with the
boundary. For the statistical process we collected and saved the final pair
of the angular variables at the end of this dynamical evolution. We see from
Fig. \ref{fig6}(a,b,c) that for $V_0<V_r$ the vast majority of points are
located inside a stability island. Such region is very close to where the
fixed point is in the vicinity of the islands. This final behaviour indeed
indicates basically the orbits are trapped inside the stability islands.
Alternatively they were attracted to their respective attracting fixed points
leading to a clear evidence that the stickiness orbits are responsible for the
decay of energy for initial velocities lower than the resonant one. To make a
comparison, in Figs. \ref{fig6}(e,f,g) we draw the histograms of frequency
of the whole evolution of the $\alpha$ variable along the dynamics for a
set of initial conditions ($10^6$), but iterated to the same number of
collisions and with the same initial velocity of the Figs. \ref{fig6}(a,b,c).
We can see some of the preferred regions for the variable $\alpha$, which
perfectly match with the position of the islands and the fixed points of the
Figs. \ref{fig6}(a,b,c). Particularly the shape of the boundaries of the
stability islands are well defined in the histogram of Fig. \ref{fig6}(g). The
same procedure is made considering now the high energy ensemble, i.e. for
$V_0>V_r$ as shown in Fig. \ref{fig6}(d). One sees there is
no longer convergence to the final pair of angular variables to the regions of
the stability islands and fixed points as seen previously. The final pair
$(\alpha,\phi)$ just stay wandering along the chaotic sea, a condition that
explains why the orbits experience Fermi acceleration. The orbits for
$V_0>V_r$ also experience stickiness, as shown in Fig. \ref{fig3}(a).
However these trappings are not sufficient condition to hold their velocity
down. Also, in Fig. \ref{fig6}(h), we make the comparison with the whole
distribution of the $\alpha$ variable along the whole dynamics, and we can see
there is no preferred region anymore. In fact the region in $\alpha$
concerning the chaotic sea shows a growth in the histogram of frequency
indicating a high chaotic behaviour therefore corroborating for the FA
phenomenon.

After a careful look at histogram of frequency we see some islands are more
preferred than others. This explains why in Fig. \ref{fig5}(a) the decay of
energy of the curves of $\omega$ produce several different plateaus of
convergence for long time which of course depend on the initial velocity. Each
of the plateaus are related to the attraction of orbits to the periodic fixed
points as shown in Fig. \ref{fig6}. The convergence plateaus are range in a
finite region of $V_{final}\in(0.55,0.15)$. In particular the curve of
$\omega$ for $V_0=0.1$ actually grows for this convergence region. Such
region seems to be the same concerning the convergence for final velocity when
dissipation is acting in the system \cite{ref35}.

\begin{figure*}[t]
\begin{center}
\centerline{\includegraphics[width=17cm,height=12.0cm]{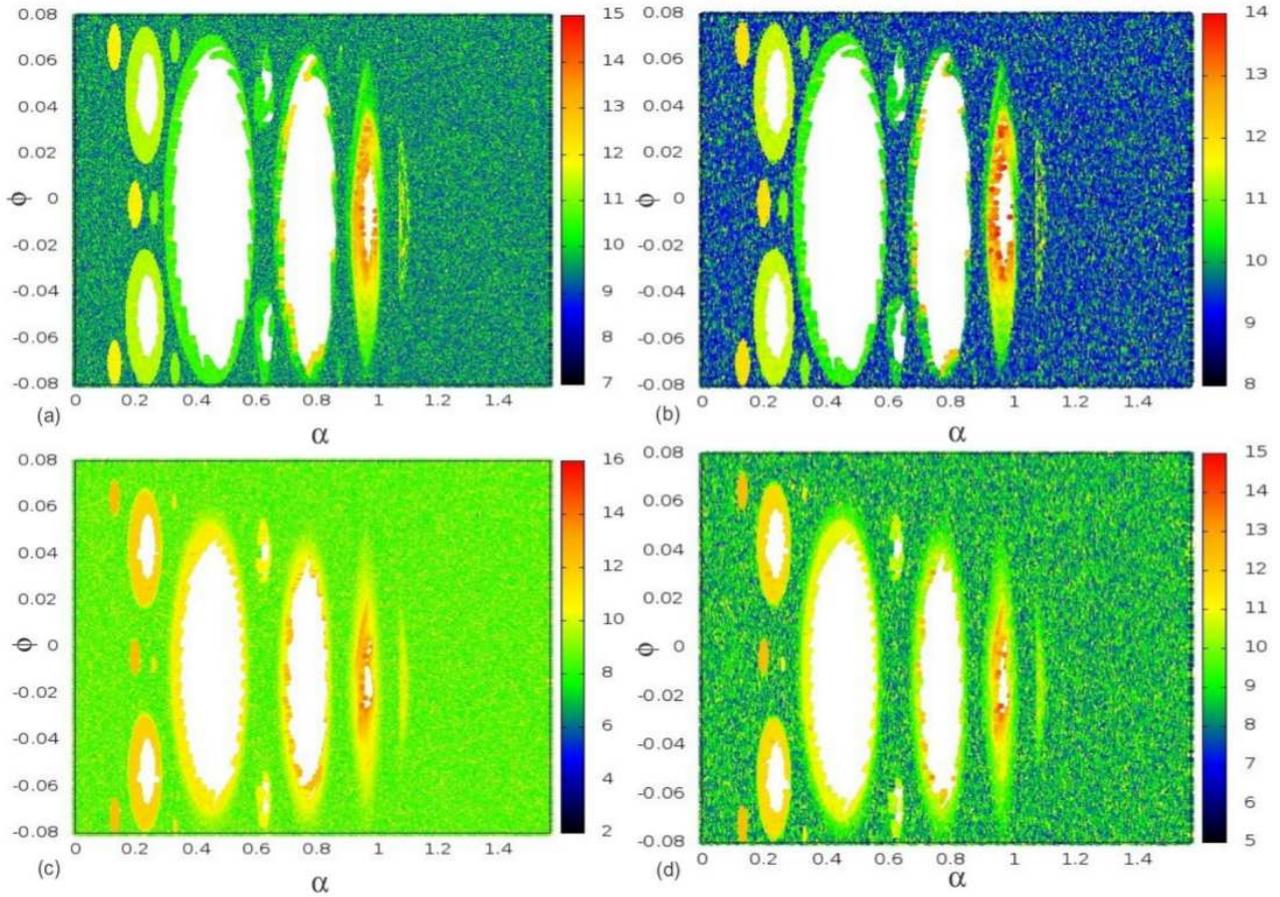}}
\end{center}
\caption{Color Online:{\it Transport analysis considering ``real escape"
orbits. In (a) and (b), we have $V_0=2.0$, and in (c) and (d) we have
$V_0=1.0$. The color scale represents the escape collision. In (a) and (c) we
have the initial conditions that crossed the resonance line and experienced
FA, and in (b) and (d), we have the initial conditions that also crossed the
resonance line, but experienced decay of energy. Note that the color
scale of all figures are in the logarithmic scale, for a better representation
of their intensity. Blue (black) indicates a fast escape, and yellow and red
(gray and dark gray) represent long times for the orbit to escape. White regions means,
that the particle never escaped.}}
\label{fig8}  
\end{figure*} 

As an attempt to quantify the different plateaus of convergence, we made a
histogram of the velocity distribution along the dynamics considering the same
ensemble as used to construct Fig. \ref{fig6} and running the dynamics until
$10^7$ collisions. Figure \ref{fig7}(a) shows these distributions for three
different values of initial velocities. One can see a major concentration for
the low energy regime between $0.1$ and $1.0$. Several peaks are noted and
decreasing in intensity as the initial velocity increases. Such behaviour is
similar as the one shown in the zoom window in Fig. \ref{fig7}(a), even when
$V_0>V_r$, which is the case of $V_0=2.0$. Also, for initial velocities lower
than the resonant one, there is a distribution for the velocity over $V_r$,
which is not so intense for $V_0=0.05$ and $V_0=0.5$ and is quite significant
for $V_0=2.0$. These distributions confirm what was supposed previously and
not all initial conditions for $V_0<V_r$ experience decay of energy. In a
complementary way, not all initial conditions with $V_0>V_r$ lead to unlimited
energy growth. Figure \ref{fig7}(b) shows the distribution of velocities
for lower energy regime as a comparison with their positions concerning the
polar angular coordinate $\alpha$. This comparison allow us to distinguish in
which island of the phase space the orbits will converge to for every range of
final velocities plateaus. For example, a comparison between Fig.
\ref{fig7}(a) and Fig. \ref{fig7}(b) shows higher intensity peaks for
velocities between $V\approx 0.29$ and $V\approx 0.35$. They indicate the
orbits prefer to stay in the last two period-one islands of the phase space
i.e., the ones located near $\alpha\approx 0.97$ and $\alpha\approx 1.10$
respectively. Finally for Fig. \ref{fig7}(c) we show the histogram of
frequency for the high energy ensemble of initial velocities. One can notice
that the several peaks are not observed anymore. Now we have a very well
defined distribution along the velocities meaning that for the very high
energy ensemble, as for example the initial velocities at least ten times
larger than $V_r$, FA is inherent in the system. Only very few orbits lead to
a decay in energy, as shown in the green curve in Fig. \ref{fig7}(c) for
$V_0=50$. We believe that the period-1 islands should be the preferred ones. 
All the results come from the resonance velocity, which marks the transition from the 
Fermi acceleration regime and the decay of energy. The analytical expression, were obtained 
considering the resonance around the period-1 islands. Also, they are the biggest ones 
in the phase space, and show themselves more influential for the stickiness phenomenon. 
You could find different resonance velocities, for different islands of different periods, 
but even so the period-1 islands should be the more influential ones. Still, one thing 
that might change the preference between the islands, would be change the value of 
the geometric control parameters. This change, would influence on the number
of islands, and also change the resonance velocity. See Refs.\cite{ref25,ref32}.

\subsection{Transport analysis}
~~~~~Let us now map along the phase space the initial conditions that lead to
unlimited growth of energy and those producing the decay of energy. To start
with we consider an ensemble of initial conditions along the phase space
uniformly distributed over $\alpha$ and $\phi$ assuming velocities either
below $V_0<V_r$ and above $V_0>V_r$ the resonant. We look at the time
evolution of each initial condition mapping then those who cross the resonant
velocity either coming from above showing a decay of energy or coming from
below therefore leading to unlimited energy growth

Considering a distribution of initial conditions equally distributed in $2000$
bins along as $\alpha\in[0,\pi/2]$ and $\phi\in[-\Phi,+\Phi]$, we evaluated
the dynamics considering an introduction a hole in the system
\cite{ref36,ref37,ref38,ref39,ref40} placed concerning the resonance velocity
$V_r=1.2$. If an initial condition started with $V_0<V_r$ achieves energy
enough to cross the resonant velocity we consider it has escaped from the low
energy region to a higher energy region. The same procedure applies for
initial conditions in the high energy regime. So if it decays to a velocity
smaller than $V_r$ we consider it has escaped from high energy to lower
energy region. We emphasize the multiple crossings from same orbit can in
principle happen so we are considering just the first crossing with $V_r$ line, up to the maximum
collision of $5\times10^6$.

Figure \ref{fig8} shows a grid of initial conditions selected near the critical region leading 
to ``escape" in a color scheme.
Figures \ref{fig8}(a,b,c,d) then represent in the color scale the respective
collision (in a log scale) where an initial condition has crossed the critical
resonance velocity line for the first time. Blue (black) indicates fast
escape, while yellow and red (gray and dark gray) denotes long time until the
orbit crosses the critical line. White regions mean that the orbits never
escaped. One can see an existence of 
orbits trapped by stickiness near the islands
concerning Fig.\ref{fig8}(a,b,c,d), indicated by yellow and red (gray and dark gray) colors.
Indeed these quasi-periodic orbits produce a delay in the diffusion along the energy/velocity 
axis and hence in the escape through the resonant velocity, leading to a delay also in Fermi
acceleration, also, they are more numerous and influential for $V_0<V_r$.

Considering first an initial velocity $V_0=2.0$, we marked
the initial conditions that escaped and reached high velocities up to the end of the simulation in
Fig.\ref{fig8}(a). In the same way the
initial conditions that had escaped and experienced the decay of energy up to the end of the simulations are drawn in Fig.
\ref{fig8}(b). The same process was made for an initial velocity $V_0=1.0$, where the initial conditions that
escaped and acquired high
velocities are drawn in Fig.\ref{fig8}(c), and finally the initial conditions
that escaped but remained at the end of the dynamics under the influence
of the decay of energy are represented in Fig.\ref{fig8}(d).

The fraction of initial conditions of $V_0=2.0$ in Figs.\ref{fig8}(a,b) is: $14.06\%$ escaped and decreased 
their velocity up to the end of simulations; $43.63\%$ escaped and experienced FA and the $42.31\%$ never 
crossed the critical line of resonance velocity. These results indicates that
the great majority of initial conditions for $V_0=2.0$ stays in the high
energy regime.
Considering now the fraction of initial conditions of $V_0=1.0$
given by Figs.\ref{fig8}(c,d) we found that: $55.41\%$ of the initial
conditions who had escaped stays in the low energy regime up to the end of
simulations, $24.04\%$ escaped and had experienced FA in the end of the
simulations, and $20.55\%$ never escaped. These results mean that for the
initial conditions of $V_0=1.0$, about $80\%$ of the orbits stay confined in
the lower energy regime. In particular, for all items of Fig.\ref{fig8}, 
one can see a strong stickiness regime in the last two islands, 
which are the islands of main influence for the convergence of orbits, as shown in 
Fig.\ref{fig7}. This strong stickiness indicates that these orbits were trapped for really long times around that 
regions, and crossed the critical resonance line in a very later time, or not even crossed it, which can be an evidence that
they were captured by the fixed points, what will lead their velocity down.\
   
\section{Final Remarks and Conclusions}
\label{sec4}
~~~~We revisited the problem of the time-dependent stadium-like billiard aiming to
understand and quantify the mechanism that is responsible for the
decay of energy. A non-linear mapping was constructed considering distinct
kinds of collisions with the boundary to describe the dynamics. A resonance
between the period of oscillation of the boundary and the rotation period
around the fixed point was confirmed. Through this resonance two ensemble
regimes can be defined at low and high energy, where for high velocities FA is
inherent.

A statistical and transport investigation along the phase space was made
concerning both regimes of initial energy as an attempt to describe the
competition between the decay of energy and FA.  We characterize the
fundamental role of stickiness and initial conditions for the existence 
of Fermi acceleration.

Focusing on the lower energy regime, we have seen that stickiness orbits lead the velocity to decay, and
for long time the dynamics is stable where most of the orbits are located very
close to the fixed points, where it seems that they were captured by fixed
points, as if the dynamics was under the influence of dissipation and then suppressing
the velocity. The results give support to new studies on stickiness
influence in the FA and diffusion process concerning systems with mixed
properties in the phase space. Also, it would be interesting to investigate, if stickiness should play similar 
role in other billiards and chaotic systems.

\section*{Acknowledgements}
ALPL acknowledges CNPq and CAPES - Programa Ci\^encias sem Fronteiras - CsF
(0287-13-0) for financial support. MPS thanks FAPESP (2012/00556-6). IBL thanks FAPESP (2011/19296-1) and EDL thanks
FAPESP (2012/23688-5), CNPq and CAPES, Brazilian agencies. ALPL and MPS also thanks the University of
Bristol for the kindly hospitality during their stay in UK. This research was
supported by resources supplied by the Center for Scientific Computing
(NCC/GridUNESP) of the S\~ao Paulo State University (UNESP).\\

\end{document}